\documentclass[12pt,showpacs,preprintnumbers,amsmath,amssymb,nofootinbib,superscriptaddress]{revtex4}
\usepackage{graphicx}
\usepackage{dcolumn}
\usepackage{bm}
\usepackage{graphics}
\usepackage{amssymb}
\usepackage{amscd}
\usepackage{afterpage}
\usepackage{float,times}
\usepackage{subfigure}
\usepackage{rotating}
\usepackage{multirow}
\usepackage{fancyheadings}
\usepackage{epsfig}
\usepackage{theorem}
\usepackage{moreverb}
\usepackage{euscript}
\usepackage{psfrag}

\begin{document}

{\hbox to\hsize{\hfill June 2007 }}


\title{Neutrino mass in radiatively-broken scale-invariant models\\}

\vspace{2 cm}

\author{Robert Foot}
  \email[Email: ]{rfoot@unimelb.edu.au}
   \affiliation{School of Physics, Research Centre for High Energy Physics,\\
    University of Melbourne, Victoria 3010, Australia.}
\author{Archil Kobakhidze}
  \email[Email: ]{archilk@unimelb.edu.au}
   \affiliation{School of Physics, Research Centre for High Energy Physics,\\
    University of Melbourne, Victoria 3010, Australia.}
\author{Kristian L. McDonald}
  \email[Email: ]{klmcd@triumf.ca}
   \affiliation{School of Physics, Research Centre for High Energy Physics,\\
    University of Melbourne, Victoria 3010, Australia.}
   \affiliation{Theory Group, TRIUMF, 4004 Wesbrook Mall, Vancouver, BC V6T2A3, Canada.}
\author{Raymond R.Volkas}
\email[Email: ]{raymondv@unimelb.edu.au}
\affiliation{School of Physics, Research Centre for High Energy Physics,\\
    University of Melbourne, Victoria 3010, Australia.}
\smallskip
 
\begin{abstract} 
\begin{center}
{\large Abstract}
\end{center}
Scale invariance may be a classical symmetry which is broken radiatively. This
provides a simple way to stabilize the scale of electroweak symmetry breaking against radiative corrections.
The simplest phenomenologically successful model of this type
involves the addition of one real scalar field to the standard model. In this minimal model the electroweak Higgs
can be interpreted as the pseudo-Goldstone boson of broken scale invariance. We study the possible origin
of neutrino mass in such models, both at tree-level and radiatively.  
\end{abstract}

\maketitle

\bigskip

\bigskip


\baselineskip=16pt
\newpage

\section{Introduction}

Scale invariance is broken explicitly in the standard model by the
$\mu^2$ Higgs parameter.
Eliminating this parameter results in a classically scale-invariant theory. However scale invariance
is anomalous: it is broken radiatively, as first studied
by Coleman and Weinberg~\cite{Coleman:1973jx}. Their analysis shows
that for a minimal scalar sector consisting of a single Higgs doublet, the scale-invariant
potential fails to produce spontaneous symmetry breaking for a top
quark in excess of about 40 GeV~\cite{Coleman:1973jx}.
The more general case with multiple scalars is therefore of interest. A suitable perturbative framework 
for analysing the general case was discussed by Gildener and Weinberg~\cite{Gildener:1976ih}.
In fact, it is quite straightforward to build phenomenologically consistent models of electroweak
symmetry breaking in scale-invariant models with multiple scalar fields~\cite{Hempfling:1996ht,Meissner:2006zh,Espinosa:2007qk,Chang:2007ki,Foot:2007as}. 

There are obvious advantages of such scale-invariant theories over the usual standard model.
In particular, they provide an elegant solution to the hierarchy problem since the electroweak
scale is then naturally stabilised with respect to radiative corrections. 

Scale invariance also forbids higher-dimensional operators, such as the
dimension-six baryon number violating
interactions and the like, which, from the
standard model point of view could exist but would spoil the renormalisability of 
the theory. In other words, scale-invariant theories provide a symmetry reason
for the non-existence of troublesome higher-dimensional terms, which is quite different
from the standard argument of imposing renormalisability. 

An important remaining issue, though, is what
to do about gravity, which of course also involves a scale, the Planck mass. It is tempting to apply scale invariance to the whole fundamental theory, including 
gravity~\cite{Foot:2007iy}, by, for example, generating the Planck mass
from the vacuum expectation value of a scalar field~\cite{Zee1979}.  For the purposes of this paper, however,
we shall set the gravity question to one side, and solve the neutrino problem first.

There is now extremely strong evidence that neutrinos oscillate due to nonzero masses and mixings.
The purpose of this paper is to address the important issue of how
neutrino mass can be generated in classically
scale-invariant theories. We investigate mechanisms which generate
neutrino masses at tree-level and others which generate it
radiatively. We find that the usual mechanisms for generating
neutrino masses have scale invariant incarnates. However only the models
which generate neutrino mass radiatively motivate the
observed suppression of the neutrino mass scale relative to the electroweak
scale.

We also find that the introduction of neutrino mass in scale invariant
models generates an accidental symmetry. In the
simplest models which induce
neutrino mass and produce a pseudo-Goldstone boson (PGB) electroweak Higgs, a real
gauge singlet scalar field is required and the
resulting symmetry is discrete. Promoting the singlet scalar
to a complex field extends the discrete symmetry to the continuous lepton
number symmetry.

The layout of this paper is as follows. In Section~\ref{sec2} we
review the realisation of the electroweak Higgs as the PGB
of anomalous scale invariance. The generation of
neutrino mass at tree level is
considered in Section~\ref{sec3} and in Section~\ref{sec4} radiatively induced neutrino
masses are studied. We conclude in Section~\ref{sec5}.

\section{Electroweak Higgs boson as PGB of broken scale invariance}
\label{sec2}

The simplest phenomenologically consistent scale invariant model \cite{Foot:2007as} involves the addition of one real
scalar field, $S$, to the minimal standard model (SM). Here we review the 
model to provide the necessary framework for
the subsequent discussions.

The most general renormalisable scale-invariant potential is
\begin{equation}
V_0(\phi,S) = \frac{\lambda_1}{2}(\phi^{\dagger}\phi)^2 + \frac{\lambda_2}{8}S^4
+ \frac{\lambda_3}{2}(\phi^{\dagger}\phi)S^2~,
\label{a1}
\end{equation}
where $\phi \sim (1,2,+1)$ under $SU(3)_c \otimes SU(2)_L \otimes
U(1)_Y$ is the SM scalar doublet.
The above potential, and the whole Lagrangian, has an accidental
discrete $Z_2$ symmetry: $S\to -S$. We parameterise the fields in unitary
gauge through
\begin{equation}
\phi= {r \over \sqrt{2}}
\bordermatrix{&\cr                &0 \cr                & \cos\omega}~,~~S=r\sin\omega
\label{a2}
\end{equation}
The potential (\ref{a1}) is then
\begin{equation}
V_0(r, \omega)=r^4\left(\frac{\lambda_1}{8}\cos^4\omega
+\frac{\lambda_2}{8}\sin^4\omega+\frac{\lambda_3}{4}\sin^2\omega\cos^2\omega  \right).
\label{a3}
\end{equation}
This tree-level potential receives quantal corrections as per the Coleman-Weinberg
analysis. We shall work in the parameter regime where the one-loop perturbative
correction $\delta V_{1-{\rm loop}}$ is sufficiently accurate.  Ideally, one would like to directly
minimise the corrected potential $V \simeq V_0 + \delta V_{1-{\rm loop}}$, but this is impossible to
do analytically.  We instead follow the approximate procedure introduced in Ref.~\cite{Gildener:1976ih}
which is valid in our weakly-coupled theory.

The idea is to first ignore the radiative
corrections and minimise the tree-level potential (\ref{a3}). 
With $r \neq 0$ but arbitrary, there are two interesting cases (ignoring
a third case that has no electroweak symmetry breaking). 

For $\lambda_3 > 0$,
\begin{equation}
\langle \sin\omega \rangle =0~,~ \langle r \rangle = \sqrt{2} \langle \phi \rangle \equiv 
v\approx 246~ {\rm GeV} \Rightarrow \langle S \rangle = 0~,
\label{a5}
\end{equation}
with
\begin{equation}
\lambda_1(\Lambda) = 0,
\label{a55}
\end{equation}
and only the electroweak Higgs develops a nonzero vacuum expectation
value (VEV). The scale $\Lambda$ is the renormalisation
point where $\lambda_1$ vanishes, anticipating the addition of radiative corrections.  The
dimensionless parameter $\lambda_1$ transmutes into the scale $\Lambda$ in the quantised theory.
This is a manifestation of the scale anomaly of quantum field theory and it generates dimensionful
quantities such as masses despite the classical scale invariance.

For $\lambda_3 < 0$, 
\begin{eqnarray}
\langle \tan^2\omega \rangle &=& \epsilon~, \nonumber \\
~\sqrt{2}\langle \phi \rangle  &=& 
\langle r \rangle \left(\frac{1}{1+\epsilon}\right)^{1/2}\equiv v\approx 246~{\rm GeV}~,~
\langle S \rangle = v\langle \tan\omega \rangle ~,
\label{a6}
\end{eqnarray}
with
\begin{equation}
\lambda_3(\Lambda)+\sqrt{\lambda_1(\Lambda)\lambda_2(\Lambda)}=0
\label{a66}
\end{equation}
where $\epsilon \equiv \sqrt{\frac{\lambda_1(\Lambda)}{\lambda_2(\Lambda)}}$. 
Once again, the required relation between the Higgs potential parameters serves
to define the renormalisation point, and a dimensionless parameter is transmuted
into the scale $\Lambda$.  In this case both scalar fields develop nonzero VEVs and the
discrete $Z_2$ symmetry is spontaneously broken.

We next calculate the tree-level Higgs masses. We first define the shifted fields
$\phi = \langle \phi \rangle + \phi'$, $S = \langle S \rangle + S'$, and substitute
them in the potential, Eq.~(\ref{a1}). 
In each case, of the two physical scalars, only one (which we shall call $H$)
gains a tree-level mass.  The other (call it $h$) is massless due to a flat direction  
in the Higgs potential. It is the PGB of anomalously-broken scale
invariance.

For the $\lambda_3 > 0 \ \Rightarrow \langle S \rangle = 0$ case,
\begin{eqnarray}
m_H^2 = {\lambda_3 v^2 \over 2} \ , \ \ H = S,
\end{eqnarray}
and the PGB is $h = \phi'_0$.

For the $\lambda_3 < 0$ case, where both $\langle S \rangle$ and $\langle \phi \rangle$ 
are nonzero, we have
\begin{eqnarray}
m_H^2 = \lambda_1 v^2 - \lambda_3 v^2 \ , \ \ 
H = -\sin\omega \phi'_0 + \cos\omega S',
\end{eqnarray}
and the PGB is $h = \cos\omega \phi'_0 + \sin\omega S'$. 

For each case, the PGB gains mass from the quantal corrections.
The $1$-loop correction along the flat direction in $V_0$
is \cite{Gildener:1976ih}
\begin{equation}
\delta V_{\rm 1-loop}= Ar^4 \ + \
Br^4\log\left(\frac{r^2}{\Lambda^2}\right )~,
\label{14}
\end{equation}
where
\begin{equation}
A=\frac{1}{64\pi^2 \langle r \rangle^4}\left[3{\rm
Tr}\left(M_V^4\log\left(\frac{M_V^2}{\langle r \rangle^2}\right)\right)+{\rm
Tr}\left(M_S^4\log\left(\frac{M_S^2}{\langle r \rangle^2}\right)\right)-4{\rm
Tr}\left(M_F^4\log\left(\frac{M_F^2}{\langle r \rangle^2}\right)\right) \right ]~,
\label{15}
\end{equation}
and 
\begin{equation}
B=\frac{1}{64\pi^2 \langle r \rangle^4}\left [3{\rm Tr}M_V^4+{\rm Tr}M_S^4-4{\rm
Tr}M_F^4
\right ]~.
\label{16}
\end{equation}
The traces go over all internal degrees of
freedom, with $M_{V,S,F}$ being the tree-level masses respectively for
vectors, scalars and fermions evaluated for the given VEV pattern.

The extremal condition $\frac{\partial \delta V_{\rm 1-loop}}{\partial
r}|_{r=\langle r \rangle}=0$ tells us that
\begin{equation}
\log\left(\frac{ \langle r \rangle}{\Lambda} \right)=-\frac{1}{4}-\frac{A}{2B}.
\label{17}
\end{equation}
The PGB mass is then \cite{Gildener:1976ih}:
\begin{eqnarray}
m_h^2  & = & \frac{\partial^2 \delta V_{\rm 1-loop}}{\partial r^2 }
|_{r = \langle r \rangle} = 8 B \langle r \rangle^2  \nonumber \\
& = & 
\frac{1}{8\pi^2 \langle r \rangle^2}\left [3{\rm Tr}M_V^4+{\rm Tr}M_S^4-4{\rm
Tr}M_F^4
\right ]~.
\label{18}
\end{eqnarray}
Applying this general formula to our theory we obtain
\begin{eqnarray}
m_h^2 &\simeq & {1 \over 8\pi^2 \langle r \rangle^2} \left[ 6m_W^4 + 3m_Z^4 + m_H^4 - 12 m_t^4\right]
\nonumber \\
& \approx & \frac{m_H^4 \cos^2\omega}{8\pi^2v^2}~,
\label{a9}
\end{eqnarray}
where the approximation follows from requiring $m_H$ to dominate 
over the other terms so as to make the PGB mass larger than the experimental 
lower limit of about $\sim 115$ GeV. For $m_h$ at this experimental lower limit the
model is consistent with precision electroweak tests provided that $\tan\omega < 0.65$, which
means that the PGB mainly ``resides'' in the electroweak doublet~\cite{Foot:2007as}.

\section{Tree-level neutrino mass generation}
\label{sec3}

We now explore two different ways that nonzero neutrino masses can be generated at tree-level
in a classically scale-invariant theory.

\subsection{Gauge singlet fermions}
If right-handed neutrinos $\nu_R$ exist, then they can Yukawa-couple to $S$,
\begin{eqnarray}
f_\nu\overline{\nu}_R \nu_R^c S + H.c..
\label{eq:nuRS}
\end{eqnarray} 
For $\langle S \rangle \neq 0$ the $\nu_R$ will gain
Majorana masses and the neutrino mass matrix becomes
\begin{eqnarray}
\mathcal{M_\nu}=
\left(\begin{array}{cc}0&f \langle \phi \rangle \\f\langle\phi\rangle&f_\nu\langle
    S\rangle\end{array}\right)
\end{eqnarray}
in the Majorana basis. The seesaw mechanism may be implemented if one
takes $f \langle \phi \rangle \ll f_{\nu} \langle S \rangle $ (see Ref.~\cite{Meissner:2006zh} for
a related model that also has this see-saw feature). 
For $f_\nu \sim 1$ and $\langle \phi \rangle \sim \langle S \rangle$, $m_{\nu_L} \sim 1$ eV 
requires $f\sim 10^{-6}$. Thus experimentally viable neutrino masses
are produced but no insight into the relative lightness
of the neutrinos is obtained. This is true also in the case where
$\langle S \rangle= 0$, where Dirac neutrinos
result and one requires $f \sim 10^{-11}$. 

Note that the accidental discrete $Z_2$ symmetry ($S \to -S$) obtained in the model without $\nu_R$ 
is now enlarged to include $\nu_R \to i \nu_R$, $\ell_L \to i\ell_L$ and $e_R \to ie_R$, 
where $\ell_L$ is the left-handed lepton doublet and $e_R$ is the right-handed charged lepton, respectively\footnote{Since the discrete charges are assigned to the standard model particles as well, 
they are defined up to hypercharge rotations. 
For instance, by hypercharge transformations we can make $\ell$ inert under the discrete symmetry, 
while the electroweak Higgs boson and quarks will transform accordingly. }.  
Variations of this accidental symmetry generically occur 
in all the models considered in this paper. 
The accidental $S\to -S$ symmetry would be violated by interactions in the scalar potential 
involving odd powers of the singlet field $S$. However classical scale 
invariance permits only quartic terms in the scalar potential,
hence one can always extend the discrete $Z_2$ of Section~\ref{sec2}
to include non-singlet scalars. This symmetry appears to be a discrete subgroup of 
the continuous lepton number symmetry, and remains a classical
symmetry of the theory. 

   Spontaneous breaking of this discrete symmetry  with $\langle S \rangle \sim \langle \phi \rangle$ leads to a potential 
cosmological domain wall problem. The standard solution is provided by inflation with a reheating temperature 
lower than the order parameter ($\approx \langle S \rangle$) of the phase transition during which the walls are produced. 
Note also that the accidental symmetries in the models considered here are anomalous and thus  the domain walls are 
not strictly stable \cite{Preskill:1991kd}. We expect that electroweak instantons and sphalerons 
(at high temperatures) will contribute to the decay of the domain
walls. A detailed study of these issues is 
beyond the scope of the present paper.
\subsection{Triplet Higgs}

Neutrino mass can also be generated by including an electroweak
triplet Higgs. In this case $\nu_R$ is not
required and neutrino masses are generated from the Lagrangian term
\begin{eqnarray}
{\cal L} = \lambda \bar \ell_L \Delta \tilde{\ell_L} + H.c..
\end{eqnarray}
Here 
\begin{equation}
\Delta \sim (1, 3, -2) = \left( \begin{array}{cc}
\Delta^-/\sqrt{2} & \Delta^0 \\ \Delta^{--} & -\Delta^-/\sqrt{2} 
\end{array} \right)
\end{equation} 
transforms like $\Delta \to U \Delta U^{\dagger}$ under $SU(2)_L$ and
$\tilde{\ell_L} \equiv i\tau_2 (\ell_L)^c \to U\tilde{\ell_L}$.
A small VEV for the
electrically neutral component $\Delta^0$ generates a tree-level Majorana mass for $\nu_L$.

The simplest phenomenologically-consistent scale-invariant potential which can give
$\langle\Delta^0\rangle \neq 0$ requires $\phi$, $\Delta$ and the real gauge singlet scalar field, 
$S$.\footnote{The minimal scale-invariant Higgs potential containing only $\phi$ and 
$\Delta$ will preserve lepton number,
and if $\langle \Delta \rangle \neq 0$ will lead to an experimentally excluded Majoron, as well
as other light scalars.}
The most general tree-level potential is then
\begin{eqnarray}
V_0  &=& \lambda_1 (\phi^{\dagger}\phi)^2 + \lambda_2 ({\rm Tr}\Delta^{\dagger}\Delta)^2 +
\lambda'_2 {\rm Tr}(\Delta^{\dagger}\Delta\Delta^{\dagger}\Delta) + 
{\lambda_3 \over 4} S^4 + \lambda_4 \phi^{\dagger}\phi {\rm Tr}\Delta^{\dagger}\Delta 
+ \lambda'_4 \phi^{\dagger}\Delta \Delta^{\dagger} \phi \nonumber \\
& + & \lambda_5 \phi^{\dagger}\phi S^2
+ \lambda_6 \Delta^{\dagger}\Delta S^2 + \lambda_7 \phi \Delta \phi S + H.c.
\end{eqnarray}
Note that only the $\lambda_7$ term violates lepton number. If in the limit $\lambda_7 \to 0$ the
parameters are such that $\langle \phi_0 \rangle = v$, $\langle S \rangle = w$, $\langle \Delta \rangle = 0$ then
taking $\lambda_7$ small but nonzero will induce a VEV for the real part of the neutral component of $\Delta$:
\begin{eqnarray}
\langle \Delta_0 \rangle = - {\lambda_7 w \over \sqrt{2}(\lambda_4 + 2\lambda_6 w^2/u^2)}.
\end{eqnarray}
Minimising the tree-level potential in the limit $\lambda_7 \to 0$ leads to the relations,
\begin{eqnarray}
\lambda_5 (\Lambda) = - \sqrt{\lambda_1 (\Lambda) \lambda_3 (\Lambda)}
\end{eqnarray}
and 
\begin{eqnarray}
{w^2 \over v^2} = \sqrt{{\lambda_1 (\Lambda) \over \lambda_3 (\Lambda)}}.
\end{eqnarray}
A small but nonzero $\lambda_7$ induces order $\lambda_7^2$ corrections to these formulas.

We can calculate the tree-level masses by expanding around the vacuum: $\phi = \langle \phi \rangle + \phi'$,
$S = \langle S \rangle + S'$ and $\Delta = \langle \Delta \rangle + \Delta'$. Taking the limit $\lambda_7 \to 0$ we find that 
the physical scalar spectrum consists of an approximately degenerate complex
$\Delta'$ triplet, a massive singlet $H = -\sin\theta \phi_0' + \cos \theta S'$, and a massless state 
$h = \cos\theta \phi_0' + \sin\theta S'$ (this is the PGB which will gain mass at one-loop level), where
\begin{eqnarray}
\tan^2 \theta &=& \sqrt{{\lambda_1 \over \lambda_3}}, \nonumber \\
m_{\Delta}^2 &=& {\lambda_4 \over 2} v^2 + \lambda_6 w^2, \nonumber \\
m_H^2 &=& 2\lambda_1 v^2 - 2\lambda_5 v^2.
\end{eqnarray}
Observe that there is no pseudo Goldstone boson associated with lepton number violation. 
This is because lepton number is explicitly broken and in the limit where the explicit lepton number violating term ($\lambda_7\phi \Delta \phi S$)
vanishes our parameter choice is such that lepton number is {\it not} spontaneously broken.

The PGB state, $h$, is massless at tree-level due to the presence of a
flat direction in the potential. This
flat direction receives one-loop corrections, leading to spontaneous symmetry breaking
as long as the bosonic degrees of freedom dominate over the top quark contribution.
As in the previous section, we end up with a relation between the PGB mass, $m_h$, and
the masses of the heavier scalars:
\begin{eqnarray}
m_h^2 \approx {1 \over 8\pi^2 (v^2 + w^2)} \left[ m_H^4 + 6m_{\Delta}^4 \right].\label{delta_mass_relation}
\end{eqnarray}
Note that in this case the PGB need not mainly `reside' in the electroweak doublet if $m_{\Delta} > m_H$. 

\section{Radiative neutrino mass generation}
\label{sec4}

Generating neutrino masses at tree-level is one interesting
possibility. Recall that both of the models considered in Section~\ref{sec3}
involved small parameters; the model with gauge singlet fermions required $f/f_\nu \sim 10^{-6}$, whilst
the model with a triplet Higgs required $\lambda_7$ similarly
small. These hierarchies are a manifestation of the
radiative breaking of scale invariance and may be understood as follows. Models which are classically
scale invariant possess no dimensional quantities in the
Lagrangian. If scale invariance is broken radiatively dimensional
transmutation occurs and a mass scale is introduced into the
theory. Identifying this with the electroweak scale means that
additional
mass scales are only present in the model if one introduces
hierarchies amongst the dimensionless parameters. If neutrinos acquire
mass at tree level such parameter hierarchies are necessary.
An alternative possibility is to generate neutrino masses radiatively. In this case the
dimensionful parameter introduced via dimensional transmutation is
only communicated to the neutrino sector by higher order processes. This renders radiatively
induced neutrino masses more appealing
from a theoretical point of view in models with anomalous
scale-invariance. In
what follows we briefly consider the simplest one-loop models, but save a more detailed analysis for the
two-loop case which has the simplest Higgs potential.

\subsection{Zee model}

The Zee model \cite{Zee:1980ai} is obtained by adding the scalar fields
\begin{eqnarray}
\phi'\sim (1,2,-1),\qquad \zeta \sim(1,1,+2)
\end{eqnarray}
to the SM (without right-handed neutrinos). 
These scalars couple to the leptons via:
\begin{eqnarray}
{\cal L} =  \lambda_1 \zeta \bar \ell_L \tilde{\ell_L} + \lambda_2 \bar \ell_L \phi' e_R + H.c.
\end{eqnarray}
where as before, $\tilde{\ell_L} \equiv i\tau_2 (\ell_L)^c$ 
and generation indices are suppressed.

\begin{figure}[b] 
\centering
\includegraphics[width=0.5\textwidth]{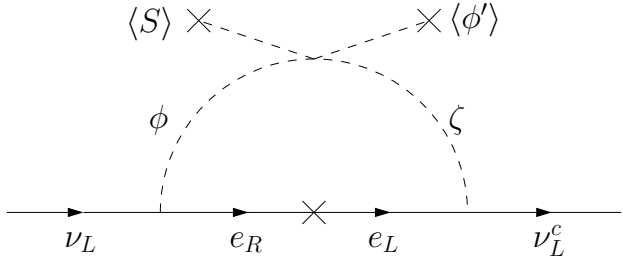} 
  \caption{Neutrino mass in a scale invariant Zee Model}
  \label{fig:zee_diagram}
\end{figure}

Enforcing scale invariance permits the electroweak Higgs to be a
PGB. However neutrino masses are not generated as scale invariance
forbids the lepton number violating term $\mu \zeta \phi\phi'$ in the
potential, ensuring that the complete scale-invariant Lagrangian conserves
lepton number. If one includes the real gauge-singlet scalar $S$ then
the electroweak Higgs remains as a PGB and the scalar potential contains the lepton
number violating term
\begin{eqnarray}
\lambda' \zeta \phi^{\dagger} \phi'S +H.c.
\end{eqnarray}
Thus provided that $\langle S\rangle \ne 0$, neutrinos acquire mass through
the one-loop diagram in Figure~\ref{fig:zee_diagram}\footnote{For $\langle
S\rangle=0$ neutrinos acquire mass through a three-loop diagram.  However, in that case the resulting masses
are smaller than the experimentally required values if the model is
required to be perturbative.}.

\subsection{A model with coloured scalars}

\begin{figure}[b] 
\centering
\includegraphics[width=0.5\textwidth]{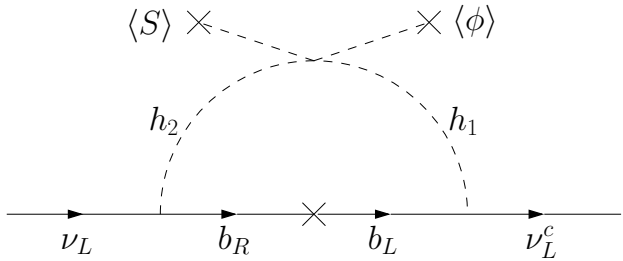} 
  \caption{Neutrino mass in a scale invariant model containing
    coloured scalar fields $h_{1,2}$.}
  \label{fig:colour_higgs_diagram}
\end{figure}

Neutrino mass may be generated radiatively if one adds coloured
scalars to the scale-invariant model outlined in Sec.~\ref{sec2}. 
Specifically, the inclusion of the scalars
\begin{eqnarray}
h_1\sim(3,1,-2/3),\qquad h_2\sim(\bar{3},2,-1/3)
\end{eqnarray}
permits the Yukawa couplings
\begin{eqnarray}
{\cal L} = \lambda_1 \bar Q_L h_1 \tilde{\ell_L} + \lambda_2 \bar \ell_L h_2 d_R + H.c.,
\end{eqnarray}
where generation indices have been suppressed.
Lepton number is violated in the scalar potential by the term
\begin{eqnarray}
S(\lambda' h_1h_2\phi + H.c.).
\end{eqnarray}
Lepton number violation is communicated to the neutrino sector
radiatively and for $\langle S\rangle\ne 0$ the one -loop diagram shown
in Figure~\ref{fig:colour_higgs_diagram} generates Majorana neutrino
masses. Scale invariance of the Lagrangian mandates that the coloured
scalars acquire electroweak-scale masses. A baryon number type symmetry can be
introduced to forbid rapid proton decay (where $h_1$ and $h_2$ have baryon charges $1/3$ and $-1/3$ respectively). 

\subsection{Babu model}

Babu's model~\cite{Babu:1988ki} is obtained by extending the SM to
include two $SU(2)_{L}$ singlet Higgs fields: a singly charged field
\mbox{$\chi^{+}$} and a doubly charged field \mbox{$k^{++}$},
\begin{eqnarray}
\chi^+ \sim(1,1,+2),\qquad k^{++} \sim(1,1,+4).
\end{eqnarray}
Thus the charged singlet in Zee's model, $\chi$, is retained whilst the
scalar doublet $\phi'$ is not. This significantly simplifies the
scalar potential. Babu's model contains the non-SM 
Yukawa couplings,
\begin{eqnarray}
\mathcal{L_{Y}} = f_{ab}\overline{(\ell_{aL})^{c}} \ell_{bL} \chi^{+} + 
h_{ab}\overline{(e_{aR})^{c}} e_{bR} k^{++}+H.c.,
\end{eqnarray}
where $a,b$ label generations, \mbox{$f_{ab}=-f_{ba}$} and \mbox{$h_{ab}=h_{ba}$}.
Gauge invariance precludes the singlet-Higgs fields from coupling to
quarks. The Higgs potential contains the terms
\begin{equation}
V(\phi, \chi^{+},
k^{++})\supset\mu \chi^{-}\chi^{-}k^{++}+\mu^*\chi^{+}\chi^{+}k^{--} + \ldots,
\label{babu_lepton_breaking}
\end{equation}
which violate lepton number by two
units. The breaking of lepton number is communicated radiatively to
the fermion sector and gives rise to Majorana neutrino masses at the two-loop
level. The resulting neutrino spectrum has been compared to the
oscillation data in~\cite{Babu:2002uu}. 

\begin{figure}[b] 
\centering
\includegraphics[width=0.5\textwidth]{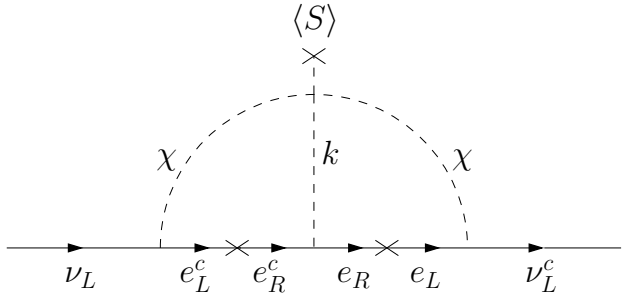} 
  \caption{Neutrino mass in a scale invariant Babu Model}
  \label{fig:babu_diagram}
\end{figure}

We would like to investigate a scale-invariant version of
Babu's model. Scale invariance forbids the lepton number violating
terms (\ref{babu_lepton_breaking}) and thus neutrinos remain massless
unless one includes the real gauge singlet field $S$. 
The complete scalar potential then contains the terms
\begin{equation}
V(\phi, \chi^{+}, k^{++},S) \supset  
S(\lambda'\chi^{-}\chi^{-}k^{++}+\lambda'^*\chi^{+}\chi^{+}k^{--}) + \ldots
\label{babu_lepton_breaking_si}
\end{equation}
which break lepton number by two units. Neutrinos again acquire mass at
the two-loop level as shown in Fig.~\ref{fig:babu_diagram}. 

These masses are
calculable and to lowest order the mass matrix takes the form
\begin{eqnarray}
M_{ab}=8\lambda'\langle S\rangle f_{ac}\tilde{h}_{cd}m_{c}m_{d}I_{cd}(f^{\dag})_{db}, 
\end{eqnarray}
where \mbox{$\tilde{h}_{ab}=\eta h_{ab}$} with \mbox{$\eta =1$} for
\mbox{$a=b$} and \mbox{$\eta =2$} for
\mbox{$a\ne b$}, $m_{c,d}$ are charged lepton masses and $I_{cd}$ is the integral
over unconstrained loop momenta~\cite{McDonald:2003zj},
\begin{eqnarray}
I_{cd}&=&\int \frac{d^{4}p}{(2\pi)^{4}}\int
\frac{d^{4}q}{(2\pi)^{4}}\frac{1}{(p^{2}-m_{\chi}^{2})}\frac{1}{(p^{2}-m_{c}^{2})}\frac{1}{(q^{2}-m_{\chi}^{2})}\frac{1}{(q^{2}-m_{d}^{2})}\frac{1}{(p-q)^{2}-m_{k}^{2}}.
\end{eqnarray}
In the limit $m_k\gg
m_\chi$ one obtains
\begin{eqnarray}
I_{cd}\simeq
\frac{1}{(4\pi)^{4}}\frac{1}{m_{k}^{2}}\left\{\log^{2}\frac{m_{\chi}^{2}}{m_{k}^{2}}
 +\frac{\pi^{2}}{3}-1\right\} \label{i_cd_large_k},
\end{eqnarray}
whilst for $m_\chi\gg m_k$ it reduces to
\begin{eqnarray}
I_{cd}\simeq\frac{1}{2^{7}\pi^{4}}\frac{1}{m_{\chi}^{2}}\left\{\frac{\pi^{2}}{6}
  +\frac{1}{2}\frac{m_{k}^{2}}{m_{\chi}^{2}}\log\frac{m_{k}^{2}}{m_{\chi}^{2}}\right\}.
\end{eqnarray}
Defining
\begin{eqnarray*}
K_{cd}=8\lambda'\langle S\rangle \tilde{h}_{cd}m_{c}m_{d}I_{cd},
\end{eqnarray*}
where no summation is implied by the repeated indices, allows one to
write the mass matrix as
$M_{ab}=(fKf^{\dagger})_{ab}$. Thus $\mathrm{Det}M=|\mathrm{Det}
 \ f|^{2}\mathrm{Det} K=0$ for an odd number of generations (due to the anti-symmetry of $f$) and to
lowest order the neutrino spectrum contains one massless state.
Let us now examine the scalar sector of the model in more detail.
The full scale-invariant Higgs potential contains eleven parameters:
\begin{eqnarray}
V &=& \lambda_1 (\phi^{\dagger}\phi)^2 + {\lambda_2 \over 4} S^4 + \lambda_3 (\chi^{\dagger} \chi)^2 + \lambda_4 (k^{\dagger} k)^2
\nonumber \\
&+& \lambda_5 \phi^{\dagger}\phi \chi^{\dagger} \chi + \lambda_6 \phi^{\dagger}\phi k^{\dagger}k + 
\lambda_7 \phi^{\dagger}\phi S^2 + \lambda_8 \chi^{\dagger} \chi k^{\dagger}k \nonumber \\
&+& \lambda_9 \chi^{\dagger} \chi S^2 + \lambda_{10} k^{\dagger} k S^2 + \lambda_{11} \chi^{\dagger} k \chi^{\dagger} S
+ H.c.
\end{eqnarray}
If all the Yukawa terms are positive except for $\lambda_7$, then the tree-level potential is minimised with
\begin{eqnarray}
\langle \phi_0 \rangle = v,\ \  \langle S \rangle = w,\ \  \langle \chi \rangle = 0,\ \  \langle k \rangle = 0
\end{eqnarray}
and
\begin{eqnarray}
\lambda_7 (\Lambda) = -\sqrt{ \lambda_2 (\Lambda) \lambda_1 (\Lambda)}, \qquad
{w^2 \over v^2} =  \sqrt{ {\lambda_1 \over \lambda_2}}.
\end{eqnarray}
As in the previous cases, we can find the masses of the physical scalars at tree-level
by expanding around the vacuum. Doing this exercise we find:
\begin{eqnarray}
m^2_k &=& {\lambda_6 \over 2}v^2 + \lambda_{10} w^2,               \nonumber \\
m^2_{\chi} &=& {\lambda_5 \over 2} v^2 + \lambda_9 w^2,                    \nonumber \\
m^2_H &=& 2\lambda_1 v^2 - 2\lambda_7 v^2\ \ {\rm where}\ \  H = -\sin\theta \phi'_0 + \cos\theta S', \nonumber \\
m^2_h &=& 0\ \ {\rm where}\ \   h = \cos\theta \phi'_0 + \sin\theta S',
\end{eqnarray}
and $\tan^2 \theta \equiv w^2/v^2$.
The state $h$ is the PGB, which gains mass at one-loop level:
\begin{eqnarray}
m_h^2 \approx {1 \over 8\pi^2 (v^2 + w^2)} \left[ m_H^4 + 2m_{k}^4 + 2m_{\chi}^4 \right].\label{babu_mass_relation}
\end{eqnarray}

\section{Conclusion}
\label{sec5}

We have analysed several different ways that neutrinos can gain nonzero masses in 
classically scale-invariant models that have the electroweak Higgs as a pseudo-Goldstone
boson.  The neutrino masses can arise either at tree-level or radiatively.  The well-known
neutrino mass generation mechanisms (see-saw, triplet-Higgs, Zee, Babu) all have
scale-invariant extensions courtesy of a gauge-singlet scalar field
$S$. These models necessarily contain scalar mass relations like those
in equations
(\ref{a9}), (\ref{delta_mass_relation})
and (\ref{babu_mass_relation}), and thus have clear experimental
signatures. However only the models which generate neutrino mass
radiatively motivate the observed suppression of the neutrino mass scale
relative to the electroweak scale. 

We note that all of the models presented contained
an accidental anomalous discrete symmetry. 
As mentioned already, spontaneous breaking of this discrete symmetry
with $\langle S \rangle \sim \langle \phi \rangle$ may lead to a potential 
cosmological domain wall problem. This problem may be avoided by
a low reheating temperature or by non-perturbative domain wall decays
 (via electroweak instantons and sphalerons). Whilst a detailed study of
these issues is 
beyond the scope of this paper, 
we note that if one promotes the scalar singlet $S$ to a complex field
the discrete symmetry is in turn promoted to the continuous lepton number
symmetry. Thus any potential domain wall problems may be
eliminated. For $\langle S \rangle\ne 0$ lepton number is
spontaneously broken and a Majoron results. Provided the Majoron is
contained mostly
within the imaginary component of $S$ this need not pose any phenomenological
concerns (see, for example,~\cite{Chang:1988aa}).

\section{Acknowledgements}

This work was supported by the Australian Research Council.

\end{document}